\documentclass[journal]{IEEEtran}
\usepackage{epsf}
\usepackage{graphicx}
\usepackage{amsmath,bm}
\usepackage{amssymb}
\usepackage{epsfig,latexsym,amsmath,epsf,amssymb,amsfonts}
\usepackage{verbatim}
\usepackage{placeins}
\usepackage[hang]{subfigure}
\usepackage{epstopdf}
\usepackage{multirow}
\usepackage{cite} 
\usepackage{amsthm}
\usepackage{url}
\usepackage{cases}
\usepackage[utf8]{inputenc}
\usepackage{chngcntr}
\usepackage{tabu}
\usepackage[linesnumbered,ruled]{algorithm2e}
\usepackage[justification=centering]{caption}
\usepackage{siunitx}

\counterwithin{paragraph}{section} 

\newcommand{\ignore}[1]{}
\usepackage{color, colortbl}

\newcommand \CV{C_{\mathrm V}}

\begin{document}
\title{\huge 3D Placement of an Unmanned Aerial Vehicle Base Station (UAV-BS) for Energy-Efficient Maximal Coverage}
\author{Mohamed Alzenad, Amr El-Keyi,
Faraj Lagum, and~Halim~Yanikomeroglu
\thanks{This work was supported by the Ministry of Higher Education and Scientific Research (MOHESR), Libya,
through the Libyan-North American Scholarship Program, and TELUS Canada.}
\thanks{The authors are with the Department of Systems and Computer Engineering, Carleton University, Ottawa, Ontario, Canada. Email: \{mohamed.alzenad, amr.elkeyi, faraj.lagum, halim\}@sce.carleton.ca, M. Alzenad is also affiliated with Sirte University, Libya, and F. Lagum with the University of Benghazi, Libya.}
}
\maketitle

\begin{abstract}
Unmanned Aerial Vehicle mounted base stations (UAV-BSs) can provide wireless services in a variety of scenarios.  In this letter, we propose an optimal placement algorithm for UAV-BSs that maximizes the number of covered users using the minimum transmit power. We decouple the UAV-BS deployment problem in the vertical and horizontal dimensions without any loss of optimality. Furthermore, we model the UAV-BS deployment in the horizontal dimension as a circle placement problem and a smallest enclosing circle problem. Simulations are conducted to evaluate the performance of the proposed method for different spatial distributions of the users.    
\end{abstract}


\section{introduction}
Unmanned aerial vehicle mounted base stations (UAV-BSs) have recently gained wide popularity as a possible solution to provide wireless connectivity in a rapid manner. UAV-BSs can assist the terrestrial cellular network in a variety of scenarios. For example, in case of terrestrial BSs failure, UAV-BSs can be rapidly deployed to meet the sudden demand for wireless services\cite{zeng2016wireless,IremMagazine,Elham2017Backhaul}. Clearly, the UAV-BS should be deployed at a location that maximizes the number of covered users.

UAVs often use batteries to power their rotors and onboard electronics\cite{Survey}. The available energy is consumed in the onboard BS and in powering the UAV platform \cite{zeng2017energy}. It was shown in \cite{Survey} that the power consumed in the BS can limit the flying time of the UAV-BS by 16\%. In this letter, we only consider the power consumed in the BS and adopt the transmission power as a measure of the energy efficient utilization of the UAV-BS. Clearly, longer battery life, and hence longer mission time, can be realized by reducing the UAV-BS transmit power. However, this reduces the coverage region of the UAV-BS and hence fewer users may be served by the UAV-BS. Thus, there is a tradeoff between the UAV-BS transmit power and its coverage region and mission time. 
 
 The work in \cite{zeng2016wireless} provides an overview of UAV-assisted communications with emphasis on the use cases, network architecture, channel characteristics, and UAV design considerations. The work in \cite{Hourani} developed a model for the probability of LoS communication between a UAV-BS and a receiver and evaluated the altitude of the UAV-BS that maximizes the coverage region. The authors in \cite{Irem} assumed the UAV-BS transmits at full power and formulated the 3D UAV-BS placement problem as a quadratically-constrained mixed integer non-linear problem. In \cite{Elham}, the authors developed a heuristic algorithm based on particle swarm optimization. The algorithm suboptimally finds the minimum number of UAV-BSs and their locations to serve a particular region. The authors in \cite{MozaffariDrone} optimized the UAV-BS altitude that results in maximum coverage region and minimum transmit power for two cases, a single UAV-BS and two UAV-BSs.  In \cite{MozaffariEfficient}, the coverage probability as a function of altitude and antenna gain is derived. Moreover, the authors in \cite{MozaffariEfficient} developed a method to deploy multiple UAV-BSs based on circle packing.

In this letter, we propose an efficient UAV-BS 3D placement method that results in maximizing the number of covered users using the minimum required transmit power. We decouple the UAV-BS placement in the vertical dimension from the horizontal dimension which simplifies the placement problem without any loss of optimality. We show that the UAV-BS placement in the horizontal dimension can be modeled as a circle placement problem and a smallest enclosing circle problem. We also evaluate the proposed method for different levels of users heterogeneity, where we show that significant power savings can be realized for highly heterogeneous scenarios.  

\section{system model}
Consider a geographical area containing a set of non-vehicular cellular users. We assume a UAV-BS is deployed, e.g., due to a congested BS or a malfunction of the infrastructure. Clearly, the UAV-BS should serve as many users as possible, which is applicable in several scenarios, e.g., offloading traffic from a congested macro cell during a temporary event such as a festival or a sports event \cite{zeng2016wireless}. Let $\mathcal U$ denote the set of users that need to be served by the UAV-BS. Let $(x_i,y_i)$ denote the location of the user $i$ in the set $\mathcal U$., and $h_{\textup{min}}$ be the minimum allowed UAV-BS altitude.

\begin{figure}
\begin{center}
\includegraphics[ height=5.5cm, width=8cm]{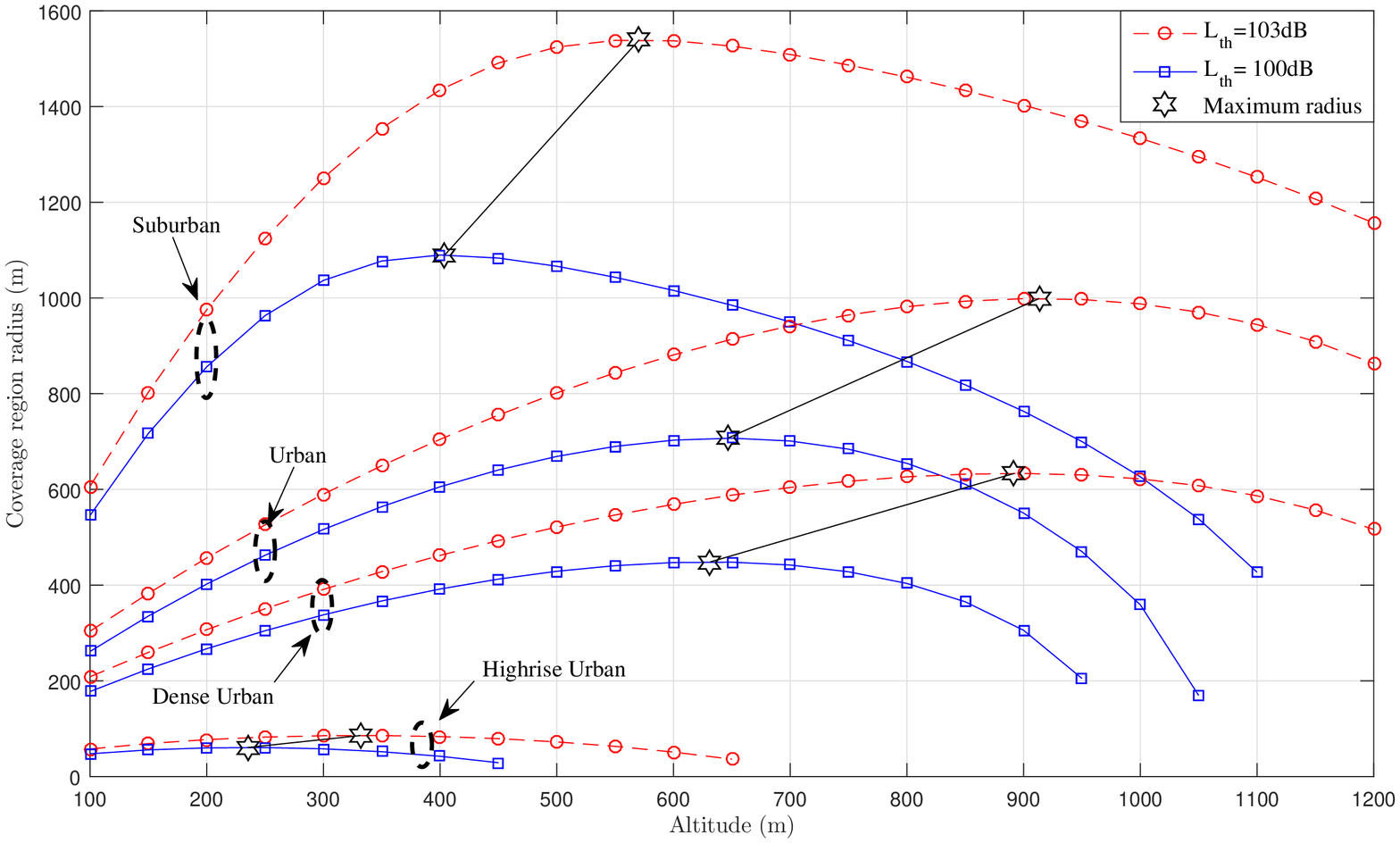}
\caption{\footnotesize Coverage region radius versus altitude for suburban, urban, dense-urban and highrise urban environments.}
\label{fig1}
\end{center}
\end{figure} 

Several models have been proposed for the air-to-ground (AtG) channels. However, we adopt the model proposed in \cite{Hourani} because of its simplicity and generality. Radio signals emitted by the UAV-BS are mainly line-of-sight (LoS) or non line-of-sight (NLoS) groups. The probability of having a LoS connection between the ground user $i$ and a UAV-BS is given by \cite{Hourani} 
\begin{equation}\label{Eq:Prob}
P_\textup{LoS}=\frac{1}{1+a \exp (-b(\frac{180}{\pi}\tan^{-1}(\frac{h}{r_i})-a))},
\end{equation} 
where $a$ and $b$ are constants that depend on the environment and are given in \cite{Hourani}, $r_i=\sqrt{(x_i-x_D)^2+(y_i-y_D)^2}$, and $(x_D,y_D)$ denotes the location of the UAV-BS in the horizontal dimension, $h$ denotes the UAV-BS altitude. Furthermore, the probability of NLoS is $P_\textup{NLoS}=1-P_\textup{LoS}$.

In addition to free space propagation loss, radio signals experience losses due to the urban environment in the form of shadowing and scattering. In this letter, we only deal with the mean path loss rather than its random behavior. This is because the planning phase of BS deployment deals with long term variations of the channel rather than small scale variations \cite{ghazzai2016optimized}. Accordingly, the path loss model for LoS and NLoS links in dB are respectively \cite{Hourani}, 
\begin{align}\label{Eq:LoSNLoSPL}
  L_\textup{LoS}&= 20\log\left(\frac{4\pi f_c d_i}{c}\right)+\eta_{\textup{LoS}} \nonumber \\
  L_\textup{NLoS}&= 20\log\left(\frac{4\pi f_c d_i}{c}\right)+\eta_{\textup{NLoS}},
\end{align}
where $f_c$ is the carrier frequency, $d_i$ is the distance between the UAV-BS and the user $i$, given by $d_i=\sqrt{h^2+r_i^2}$. Furthermore, $\eta_{\textup{LoS}}$ and $\eta_{\textup{NLoS}}$ are the mean additional losses for LoS and NLoS, respevtively, and are given in \cite{Hourani}. In the absence of terrain knowledge, we can not determine whether the link is LoS or NLoS. Therefore, we consider the probabilistic mean path loss, which is averaged over the LoS and NLoS conditions as
\begin{equation}\label{Eq:PL1}
L(h,r_i)=L_\textup{LoS}\times P_\textup{LoS}+L_\textup{NLoS}\times P_\textup{NLoS}.
\end{equation}

Let $A=\eta_{\textup{LoS}}-\eta_{\textup{NLoS}}$, $B=20\log(\frac{4\pi f_c}{c})+\eta_{\textup{NLoS}}$ and $\theta_i=\tan^{-1}(\frac{h}{r_i})$ in radians. Substituting (\ref{Eq:Prob}) and (\ref{Eq:LoSNLoSPL}) into (\ref{Eq:PL1}) and noting that $h^2+r_i^2=(\frac{r_i}{\cos( \theta_i)})^2$ yields

\begin{equation}\label{Eq:PL2}
\resizebox{.89\columnwidth}{!}{$\displaystyle{L(h,r_i)=\frac{A}{1+a \exp (-b(\frac{180}{\pi}\theta_i-a))}+20\log(\frac{r_i}{\cos (\theta_i)})+B}$}.
\end{equation}

For a given transmit power $P_t$, the received power at the user $i$ depends on the path loss experienced by its communication link, and can be written as
\begin{equation}\label{Eq:PowRec}
P_r=P_t-L(h,r_i).
\end{equation}

 In order to have a guaranteed quality of service (QoS), we assume that the received power $P_r$ must exceed a certain threshold $P_\textup{min}$. This is equivalent to saying that the user $i$ is covered if its link experiences a path loss less than or equal to a certain threshold $L_\textup{th}$, i.e., $L(h,r_i)\leq L_\textup{th}$\footnote{The coverage is often defined in terms of the received signal-to-noise ratio (SNR). Since we implicitly assume that the noise power does not change in any significant way over time, the path loss-based coverage is identical to the SNR-based coverage.}. Mathematically, the radius of the coverage region can be defined as $R=r |_{L(h,r)= L_\textup{th}}$.

\section{deployment of UAV-BS for maximum covered users and minimum transmit power}
\subsection{Finding the altitude for maximum coverage region}
As easily seen from (\ref{Eq:PL2}), for a particular environment and a given UAV-BS altitude $h$, all the ground points, which are located at a distance $R$, experience the same path loss $L_\textup{th}$. Moreover, all points located at a distance $r\leq R$ experience a path loss smaller than $L_\textup{th}$. This is equivalent to saying that the coverage region of the UAV-BS is a circular disk. Fig.~\ref{fig1} shows the coverage region radius as a function of $h$ for different environments and for two QoS requirements i.e., $L_\textup{th}=100$ dB and $L_\textup{th}= 103$ dB. It was proven in \cite{MozaffariDrone}, which can also be seen in Fig.~\ref{fig1}, that for a given path loss, the coverage region radius $R$ as a function of $h$, has only one stationary point that corresponds to the maximum coverage region radius. Since $R$ is an implicit function of $h$, we find the unique stationary point numerically. The optimal altitude that results in the maximum coverage region can be found by solving 
\begin{equation}\label{Eq:partial}
 \frac{\partial R}{\partial h}=\frac{\partial R}{\partial \theta}\frac{\partial \theta}{\partial h}=0.
\end{equation}
Since $\frac{\partial \theta}{\partial h}=\frac{\partial }{\partial h}\tan^{-1}{(\frac{h}{r})}=\frac{r}{h^2+r^2}> 0$, searching for $h$ that maximizes $R$ can be achieved by searching for $\theta$, denoted by $\theta_{\textup{opt}}$, that maximizes $R$, i.e., solving $\frac{\partial R}{\partial \theta}=0$ for $\theta$, which yields the following equation \cite{Hourani}:
\begin{equation}\label{Eq:Theta}
\frac{\pi}{9\ln(10)}\tan{\theta_{\textup{opt}}}+\frac{abA\exp(-b(\frac{180}{\pi}\theta_{\textup{opt}}-a))}{(a\exp(-b(\frac{180}{\pi}\theta_{\textup{opt}}-a))+1)^2}=0.
\end{equation}

Let $h_1$ denote the altitude that maximizes the area of the coverage region and $R_1$ be the corresponding radius for a given path loss $L_{\textup{th}_1}$. Clearly, one can observe from (\ref{Eq:Theta}) and also the straight line in Fig.~\ref{fig1}, which has a slope of $\frac{R_1}{h_1}$, that the optimal elevation angle $\theta_{\textup{opt}}$ depends only on the environment. Solving (\ref{Eq:Theta}) numerically yields $\theta_{\textup{opt}}= 20.34^\circ,  42.44^\circ, 54.62^\circ$, and $75.52^\circ$ for the suburban, urban, dense urban and high-rise urban environments.

Given the optimal elevation angle $\theta_{\textup{opt}}$, and the path loss threshold $L_{\textup{th}_1}$, the maximum coverage radius  $R_1$ can be evaluated by solving the following equation for $R_1$ as
\begin{equation}\label{Eq:PL3}
\resizebox{.89\columnwidth}{!}{$\displaystyle{L_{\textup{th}_1}=\frac{A}{1+a \exp (-b(\frac{180}{\pi}\theta_{\textup{opt}}-a))}
+20\log(\frac{R_1}{\cos \theta_{\textup{opt}}})+B}$}.
\end{equation}
Finally, the altitude $h_1$ that maximizes the area of the coverage region is given by
\begin{equation}\label{Eq:h1OptAlt}
h_1=R_1\tan(\theta_{\textup{opt}}).
\end{equation}

\subsection{Finding the optimal 2D placement}
Since the coverage region of a UAV-BS is a circular disc, placing the circular disc on the horizontal plane corresponds to placing the UAV-BS horizontally. Let $\textup C_1$ denote the coverage region with radius $R_1$. Now, we need to optimally center the coverage region $\textup C_1$ on the horizontal plane such that it encloses the maximum possible number of users, which is known as a circle placement problem.

Let $u_i\in \left\{0,1 \right\}$ be a binary decision variable such that $u_i=1$ when the user $i$ is inside the coverage region $\textup C_1$, and $u_i=0$ otherwise. The user $i$ is covered if it is located within a distance at most $R_1$ from the center of the coverage region $\textup C_1$. This condition can be written as 
\begin{equation}\label{Prb1:Constraint1Before}
u_i((x_i-x_D)^2+(y_i-y_D)^2)\leq R^2_1.
\end{equation}
To enforce the requirement that constraint (\ref{Prb1:Constraint1Before}) be satisfied when $u_i=0$, we further rewrite constraint (\ref{Prb1:Constraint1Before}) as 
\begin{equation}\label{Prb1:Constraint1After}
(x_i-x_D)^2+(y_i-y_D)^2\leq R^2_1+M(1-u_i),
\end{equation}
where $M$ is a large constant which satisfies constraint (\ref{Prb1:Constraint1Before}) when $u_i=0$. We can see that when $u_i=1$, the constraint (\ref{Prb1:Constraint1After}) reduces to the constraint (\ref{Prb1:Constraint1Before}). On the other hand, when $u_i=0$, any choice for $(x_D,y_D)$ within the allowable deployment region will satisfy the constraint (11). The placement problem in the horizontal dimension is then formulated as
\begin{equation} \label{Prb:Disc1}
\begin{aligned} 
&\underset{x_D,y_D,{u_i}}{\operatorname{maximize}}  \hspace{0.2cm}   \sum_{i\in \mathcal U}u_i \\
&\text{subject to} \\ 
&(x_i-x_D)^2+(y_i-y_D)^2\leq R^2_1+M(1-u_i),\hspace{.08cm} \forall i\in \mathcal U \\
&u_i\in \left\{0,1\right\}, \hspace{4.94cm} \forall i\in \mathcal U. 
 \end{aligned}
\end{equation} 

Let $(x_D^1,y_D^1)$ and  $\mathcal U^\textup{ cov}\subseteq \mathcal U$ denote the optimal horizontal location of the center of the coverage region $\textup C_1$ and the set of the covered users obtained by solving (\ref{Prb:Disc1}), respectively. The problem (\ref{Prb:Disc1}) is a mixed integer non-linear problem (MINLP), which can be solved by the MOSEK solver. Obviously, the transmit power of the UAV-BS should be utilized as efficient as possible. This can be achieved if the QoS is guaranteed for the same number of covered users $\left\vert{\mathcal U^\textup{cov}}\right\vert $ with less transmit power. Now as there might be no users right on the border of the coverage region $\textup C_1$, the transmit power can be reduced to the level at which the same set of users are still covered. For a given 3D location of the UAV-BS $(h_1,x_D^1,y_D^1)$ , the reduction in transmit power results in reducing the radius of the coverage region. Clearly, the transmit power can be further reduced if the horizontal location of the UAV-BS $(x_D^1,y_D^1)$ is readjusted  which corresponds to recentring the coverage region $\textup C_1$. Therefore, we resize and recenter the coverage region $\textup C_1$ with the same set of users enclosed, which is known as a smallest enclosing circle problem. This can be achieved by solving the problem,
\begin{equation} \label{Prb:Disc2}
\begin{aligned} 
&\underset{x_D,y_D,r}{\operatorname{minimize}}  \hspace{0.2cm}   r \\
&\text{subject to} \\
&\hspace{0.1cm}(x_i-x_D)^2+(y_i-y_D)^2\leq r^2, \hspace{1.4cm} \forall i=\in \mathcal U^\textup{cov}.
  \end{aligned}
\end{equation} 

The problem (\ref{Prb:Disc2}) can be efficiently solved by transforming it to a second order cone problem (SOCP). In this letter, we use the CVX parser/solver to solve the problem (\ref{Prb:Disc2}). Let $(x_D^*,y_D^*)$ and $R_2$ denote the center and the radius of the coverage region $\textup C_2$ obtained by solving (\ref{Prb:Disc2}), respectively. Obviously, the users located right on the border of the coverage region $\textup C_2$, known as cell border users, have the highest path loss compared to the other covered users. Therefore, ensuring cell border users are within the coverage region guarantees that all other users are also covered. Clearly, these cell border users are not at the optimal elevation angle $\theta_{\textup{opt}}$. Therefore, there is a vertical location $h^{\star}\leq h_1$ that results in a reduction in the path loss experienced by all covered users. The optimal altitude that leads to maximizing the number of covered users and minimizing the transmit power is then $h^\star=\textup{max} (h_\textup{min},R_2\tan(\theta_{\textup{opt}}))$.
 
Let $L(h^{\star},R_2)$ denote the path loss experienced by cell border users located at a distance $R_2$ from the center of the coverage $\textup C_2$	when the UAV-BS is placed at the altitude $h^\star$. The minimum required transmit power is then
\begin{equation}
P_\textup {req}=P_{\textup{min}}+L(h^{\star},R_2).
\end{equation} 
The proposed algorithm is given below.
\begin{algorithm}
    \SetKwInOut{Input}{Input}
    \SetKwInOut{Output}{Output}
  \Input{$a,b,\eta_\textup{LOS},\eta_\textup{NLOS},(x_i,y_i),\theta_\textup{opt},h_\textup{min},P_t, P_\textup{min}$}
    \Output{$(x_D^\star,y_D^\star,h^\star)$}
Obtain $R_1$ by solving (\ref{Eq:PL3})

Obtain $h_1$ by solving (\ref{Eq:h1OptAlt})

\textbf{Solve a problem:} Obtain $(x_D^1,y_D^1)$ and $\mathcal U^\textup{cov}$ by solving problem (\ref{Prb:Disc1})

\textbf{Solve a problem:} Recenter and resize coverage region $\textup C_1$ to obtain $(x_D^\star,y_D^\star)$ and $R_2$ by solving (\ref{Prb:Disc2})

\textbf{Altitude recalculation:} Obtain the optimum altitude $h^\star=\textup{max} (h_\textup {min},R_2\tan(\theta_{\textup{opt}}))$

\caption{Obtain optimal 3D location $(x_D^\star,y_D^\star,h^\star)$}
\end{algorithm}
 
\section{simulation results} 
In our simulations, we consider 3 km $\times$ 3 km area. We assume that $f_c=2$ GHz, $P_t=30$ dBm, $P_\textup{min}=-70$ dBm, and $h_\textup{min}=100$ m. The users are dropped according to the Thomas point process for two users densities $\lambda_1= \SI{6}{\textup{users}/km^{2}}$ and $\lambda_2=\SI{9}{\textup{users}/km^{2}}$. For comparison, we assume a UAV-BS transmits at full power $P_t=30$ dBm and is vertically placed at the altitude that maximizes the coverage region and randomly deployed in the horizontal dimension. To measure the heterogeneity of the user distribution, we use the Coefficient of Variation (CoV) of the Voronoi cell area proposed in \cite{Mirahsan1015}. The CoV is defined as $\CV=\frac{1}{0.529} \cdot \frac{\sigma_{\mathrm V}}{\mu_{\mathrm V}}$, where $\sigma_{\mathrm V}$ and  $\mu_{\mathrm V}$ are the standard deviation and the mean of the areas of the Voronoi tessellations, respectively. $\CV=1$ corresponds to the situation where users are uniformly distributed (Poisson Point Process) while $\CV > 1$ represents the situation where users form clusters located around hotspots  ~\cite{Mirahsan1015}.

\begin{figure}
\begin{center}
\includegraphics[ height=4.8cm, width=9cm]{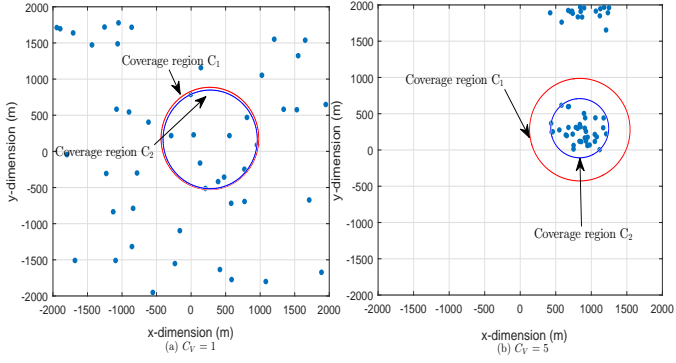}
\caption{\footnotesize Two possible 2D deployments with the same number of covered users.}
\label{fig2}
\end{center}
\end{figure}

Fig.~\ref{fig2} shows two possible user distributions for $\CV=1$ and $\CV=5$, each with two possible coverage regions $\textup C_1$ and $\textup C_2$ that maximize the number of covered users. Clearly, for a given distribution and the same UAV-BS altitude, the coverage region $\textup C_2$ is more power efficient because the UAV-BS covers the same users with less transmit power. Moreover, more power saving is realized as the users heterogeneity increases.

\begin{figure}
\begin{center}
\captionsetup{justification=centering}
\includegraphics[ height=5.5cm, width=8cm]{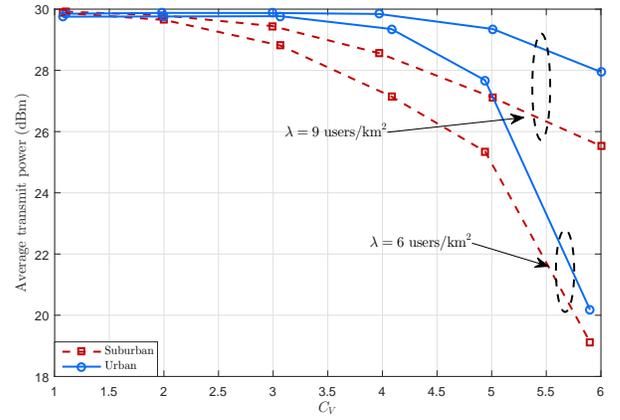}
\caption{\footnotesize Average required transmit power versus CoV.}
\label{fig3}
\end{center}
\end{figure}

Fig.~\ref{fig3} shows the average transmit power versus the CoV. As expected, for low heterogeneity scenario, the power saving is low. This is because the users are spread over the entire area and only a small area near the cell edge may not have users. In such a situation, we have $R_2\approx R_1$. However, with the increase of user heterogeneity, the average transmit power decreases for the two environments. This is due to the fact that as the user heterogeneity increases, more users become closer to each other, forming clusters that can be covered by a lower transmit power. In such a situation, we have $R_2\ll R_1$.

\begin{figure}
\begin{center}
\includegraphics[ height=5.5cm, width=8cm]{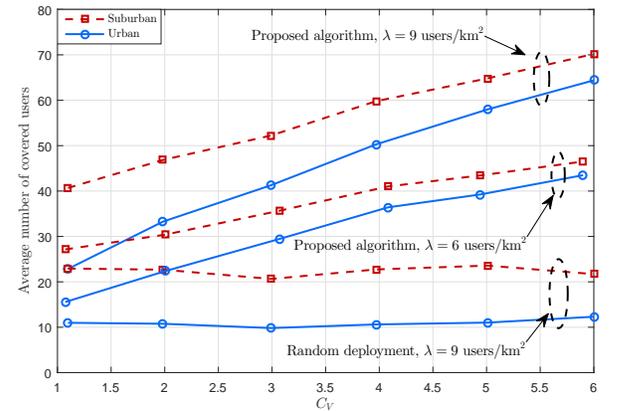}
\caption{\footnotesize Average number of covered users versus CoV.}
\label{fig4}
\end{center}
\end{figure}

Fig.~\ref{fig4} illustrates the average number of covered users versus the CoV. Clearly, more users are covered in the suburban environment than in the urban environment because the former has a wider coverage region. Moreover, for the proposed algorithm, more users are covered as the CoV increases. This is because at high CoV, users are located closer to each other and hence more users fall within the coverage region. It can also be seen that the proposed algorithm performs better than the randomly deployed UAV-BS over the entire range of the CoV, e.g., at $\CV=6$, the proposed algorithm results in covering 70 users using $P_t=25.5$ dBm while the random deployment algorithm covers 22 users using $P_t=30$ dBm.   

\section{Conclusion}
In this letter, we have studied energy-efficient 3D placement of a UAV-BS that maximizes the number of covered users using the minimum required transmit power. We have decoupled the UAV-BS placement in the vertical dimension from the horizontal dimension which simplifies the placement problem. Simulation results have shown significant savings in transmit power and increase in the number of covered users as the user heterogeneity increases.   
  
\bibliographystyle{IEEEtran}
\bibliography{IEEEfull,RefList}
\end{document}